\documentclass[%
 reprint,
superscriptaddress,
amsmath,amssymb,
aps,
prx,
]{revtex4-2}

\usepackage{graphicx}%
\usepackage{amsmath,amssymb,amsfonts}%
\usepackage{xcolor}%
\usepackage{mathtools}
\usepackage{braket}
\usepackage{array}
\usepackage{url}
\usepackage{mathrsfs}

\newcommand{\pdfrac}[2]{\frac{\partial #1}{\partial #2}}
\newcommand{\pdhfrac}[2]{\partial #1/\partial #2}

\DeclarePairedDelimiter\norm{\lVert}{\rVert}%

\newcommand{\Csca}{C_\text{sca}}
\newcommand{\Cext}{C_\text{ext}}
\newcommand{\Cpr}[1]{C_{pr,#1}}
\newcommand{\area}{L'}

\newcommand{\Lframe}{$\mathscr{L}$}
\newcommand{\Mframe}{$\mathscr{M}$}
\newcommand{\Lbasis}[1]{\hat{\mathbf{f}}_{#1}}
\newcommand{\Mbasis}[1]{\hat{\mathbf{f}}_{#1'}}
\newcommand{\Lfourbasis}[1]{\vec{f}_{#1}}
\newcommand{\Mfourbasis}[1]{\vec{f}_{#1'}}

\newcommand{\pdCprpdtheta}[1]{\pdfrac{\Cpr{#1}'}{\theta'}}

\RequirePackage{xparse}
\NewDocumentCommand\lamfrac{s}{%
  \IfBooleanTF#1%
    {\frac{\lambda}{\Lambda'D}}% If a star is seen
    {\frac{\lambda'}{\Lambda'}}%     If no star is seen
}
\NewDocumentCommand\lamtfrac{s}{%
  \IfBooleanTF#1%
    {\tfrac{\lambda}{\Lambda'D}}% If a star is seen
    {\tfrac{\lambda'}{\Lambda'}}%     If no star is seen
}
\NewDocumentCommand\lamhfrac{s}{%
  \IfBooleanTF#1%
    {\lambda/(\Lambda'D)}% If a star is seen
    {\lambda'/\Lambda'}%     If no star is seen
}
\NewDocumentCommand\Lamfrac{s}{%
  \IfBooleanTF#1%
    {\frac{\Lambda'D}{\lambda}}% If a star is seen
    {\frac{\Lambda'}{\lambda'}}%     If no star is seen
}
\NewDocumentCommand\sqrtlamfrac{s}{%
  \IfBooleanTF#1%
    {\sqrt{1-\Big(\lamfrac*\Big)^2}}% If a star is seen
    {\sqrt{1-\Big(\lamfrac\Big)^2}}%     If no star is seen
}
\NewDocumentCommand\sqrtlamtfrac{s}{%
  \IfBooleanTF#1%
    {\sqrt{1-(\lamtfrac*)^2}}% If a star is seen
    {\sqrt{1-(\lamtfrac)^2}}%     If no star is seen
}
\NewDocumentCommand\sqrtlamhfrac{s}{%
  \IfBooleanTF#1%
    {\sqrt{1-(\lamhfrac*)^2}}% If a star is seen
    {\sqrt{1-(\lamhfrac)^2}}%     If no star is seen
}

\NewDocumentCommand\mlamfrac{s}{%
  \IfBooleanTF#1%
    {\frac{m\lambda}{\Lambda'D}}% If a star is seen
    {\frac{m\lambda'}{\Lambda'}}%     If no star is seen
}
\NewDocumentCommand\mlamtfrac{s}{%
  \IfBooleanTF#1%
    {\tfrac{m\lambda}{\Lambda'D}}% If a star is seen
    {\tfrac{m\lambda'}{\Lambda'}}%     If no star is seen
}
\NewDocumentCommand\mlamhfrac{s}{%
  \IfBooleanTF#1%
    {m\lambda/(\Lambda'D)}% If a star is seen
    {m\lambda'/\Lambda'}%     If no star is seen
}
\NewDocumentCommand\sqrtmlamfrac{s}{%
  \IfBooleanTF#1%
    {\sqrt{1-\Big(\mlamfrac*\Big)^2}}% If a star is seen
    {\sqrt{1-\Big(\mlamfrac\Big)^2}}%     If no star is seen
}
\NewDocumentCommand\sqrtmlamtfrac{s}{%
  \IfBooleanTF#1%
    {\sqrt{1-(\mlamtfrac*)^2}}% If a star is seen
    {\sqrt{1-(\mlamtfrac)^2}}%     If no star is seen
}
\NewDocumentCommand\sqrtmlamhfrac{s}{%
  \IfBooleanTF#1%
    {\sqrt{1-(\mlamhfrac*)^2}}% If a star is seen
    {\sqrt{1-(\mlamhfrac)^2}}%     If no star is seen
}

\newcommand{\effr}[1]{r_{#1}'}
\newcommand{\efft}[1]{t_{#1}'}
\newcommand{\effrL}[1]{r_{#1}^{L\prime}}
\newcommand{\effrR}[1]{r_{#1}^{R\prime}}

\newcommand{\FOM}{F_\text{dmp}}

\newcommand{\Dinv}{\biggl(\frac{1}{D} - 1\biggr)}

\begin{document}

\preprint{APS/123-QED}

\title{All-optical damping forces enhanced by metasurfaces for stable relativistic lightsail propulsion}

\author{Jadon Y. Lin}
\affiliation{%
 School of Physics, The University of Sydney, Sydney, 2006, NSW, Australia
}%
\affiliation{%
Institute of Photonics and Optical Science, The University of Sydney, Sydney, 2006, NSW, Australia
}%

\author{C. Martijn de Sterke}%
\affiliation{%
 School of Physics, The University of Sydney, Sydney, 2006, NSW, Australia
}%
\affiliation{%
 Institute of Photonics and Optical Science, The University of Sydney, Sydney, 2006, NSW, Australia
}%

\author{Michael S. Wheatland}
\affiliation{%
 School of Physics, The University of Sydney, Sydney, 2006, NSW, Australia
}%

\author{Alex Y. Song}
\affiliation{School of Electrical and Computer Engineering, The University of Sydney, Sydney, 2006, NSW, Australia}
\affiliation{%
 Institute of Photonics and Optical Science, The University of Sydney, Sydney, 2006, NSW, Australia
}%
\affiliation{The University of Sydney Nano Institute, The University of Sydney, Sydney, 2006, NSW, Australia}

\author{Boris T. Kuhlmey}
\email{boris.kuhlmey@sydney.edu.au}
\affiliation{%
 School of Physics, The University of Sydney, Sydney, 2006, NSW, Australia
}%
\affiliation{%
 Institute of Photonics and Optical Science, The University of Sydney, Sydney, 2006, NSW, Australia
}%
\affiliation{The University of Sydney Nano Institute, The University of Sydney, Sydney, 2006, NSW, Australia}

\date{\today}% It is always \today, today,

\begin{abstract}
Lightsails are a promising spacecraft concept that can reach relativistic speeds via propulsion by laser light, allowing travel to nearby stars within a human lifetime. The success of a lightsail mission requires that any motion in the plane transverse to the propagation direction is bounded and damped for the entire acceleration phase. Here, we demonstrate that a previously unappreciated relativistic force, which generalizes the Poynting-Robertson effect, can passively damp this transverse motion. We show that this purely optical effect can be enhanced by two orders of magnitude compared to plane mirror sails by judicious design of the scattering response. We thus demonstrate that exploiting relativistic effects may be a practical means to control the motion of lightsails.
\end{abstract}

\maketitle

\section{\label{sec:intro}Introduction\protect}

One of the most promising approaches to spacecraft capable of traversing interstellar distances are lightsails~\cite{lubin2016,kulkarni2018,starshot2023}, crafts with extremely low mass ($\sim 1$~g) which are accelerated by reflecting photons emitted by an Earth-based high-power laser. For such crafts to reach the near-relativistic ($\approx 0.2c$) speeds necessary for travel to, say, the nearest star to the Sun, Proxima Centauri, within a human lifetime, the high-power source must stay focused on the sail for as large a distance as possible. Such a stringent requirement necessitates the use of a large (kilometer) scale array of lasers whose $\approx 50$~GW beam converges on the diffraction-limited spot size of the sail as it accelerates. However, the laser beam divergence rapidly reduces the efficiency of propulsion, and so the laser must eventually turn off to mitigate the massive operating cost~\cite{Parkin2018}. 
For a target velocity of $v=0.2c$, the acceleration phase during which the laser is turned on is estimated to last $\sim 10$~minutes with a flight distance $\sim 0.1$~AU~\cite{lubin2016}. Among the numerous technological and design challenges that must be overcome to support the sail during this acceleration phase, such as thermal management~\cite{atwater2018,Jin2020,Jin2022} and laser design~\cite{Bandutunga2021}, one of the most critical is the stability of the sail in the plane transverse to the longitudinal beam propagation direction: the lightsail must remain within the laser beam, or else it will stop being propelled. This requires mechanical asymptotic stability~\cite{Szidarovsky2017}: displacements within the beam and out-of-plane rotations need not only to be bounded by restoring forces and torques, but any resulting oscillations need to be damped. Damping is crucial to ensure that the sail is robust to perturbations in both position and velocity that are acquired throughout the laser acceleration period, for example due to imperfect beam tracking or atmospheric beam distortions. Damping forces do not come naturally in the vacuum of space and active stabilization mechanisms (\textit{e.g.} thrusters) cannot be added to the sail without exceeding the low mass budget. Furthermore, light travel times between lightsail and laser are likely to exceed the characteristic time of transverse oscillations, so damping by modulating the laser using feedback loops or parametric damping seem improbable during the acceleration phase~\cite{chu2019,Chu2021}. Therefore, damping needs to be provided passively, as an intrinsic design feature of the sail itself. 

\begin{figure*}[htb]
    \centering
    \includegraphics[width=0.7\textwidth]{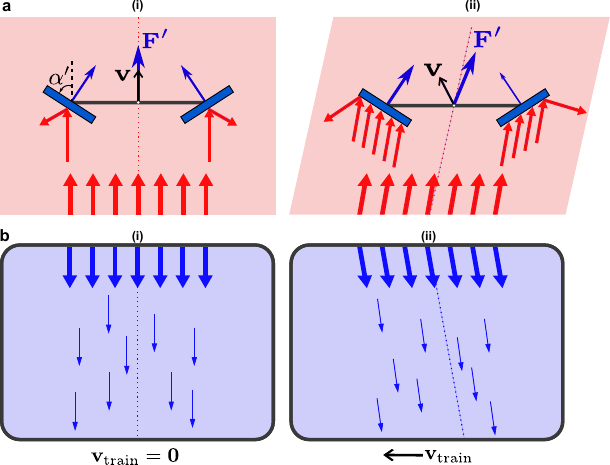}
    \caption{(a) V-mirror sail as viewed in its comoving frame. Red arrows symbolize light direction. Panel (i) shows the desired lightsail motion, whereas (ii) shows the relativistic light aberration and resultant force due to a non-zero velocity in the horizontal plane. (b) Relativistic aberration can be understood through the classical rain analogy, whereby vertical raindrops when the train is stationary, as in (i), appear slanted when the train is moving, as in (ii).}
    \label{fig:VM_panel}
\end{figure*}
To date, almost all of the reported sail designs generate restoring forces without damping forces. The restoring forces come from either: judiciously chosen, mirror-like sail geometries such as cones~\cite{singh2000,schamiloglu2001,abdallah2003,gao2024} and spheres~\cite{manchester2017}, or nanostructured reflections in diffraction gratings~\cite{ilic2019,Srivastava2019,srivastava2020,gao2024} and metasurfaces~\cite{salary2020,salary2021,Taghavi202204,Taghavi202211}. 

In contrast, designs providing the necessary damping force have been very limited: Rafat {\em et al.}~\cite{rafat2022} showed that it can be achieved using damped internal degrees of freedom, but at the cost of imposing substantial sail constraints which are likely difficult to achieve within the lightsail's mass budget. 
More recently, a geometric optics-based analysis of an idealized V-shaped mirror (Fig.~\ref{fig:VM_panel}(a)) showed that relativistic corrections lead to velocity-dependent  optical forces that can in principle provide some level of damping~\cite{Mackintosh2024}.
The origin of this drag force is similar to that of the Poynting-Robertson effect~\cite{Poynting1904,Robertson1937}, which causes small dust grains orbiting a star to lose angular momentum and to slowly spiral into their star. The effect is essentially due to the relativistic aberration of light~\cite{Einstein1905}, in which light emitted by a source and viewed by an observer moving relative to the source appears to come from an angle which varies with the velocity of the observer. The apparent shift in direction of incidence is such that the light opposes the velocity of the observer, analogous to the classical effect whereby an observer riding on a train in rainy weather sees rain drops falling with a non-zero horizontal component in the direction opposite the train's motion (depicted in Fig.~\ref{fig:VM_panel}(b)). The faster the train, the more severely the rain drops oppose the train's motion. In the case of lightsails, the sail ``observes'' the laser's photons aberrated depending on the sail's transverse velocity, with a faster sail resulting in greater aberration. For instance, the sail with a leftward transverse velocity as in Fig.~\ref{fig:VM_panel}(a,ii) experiences the plane wave from the laser approach from the left as opposed to the usual incidence in Fig.~\ref{fig:VM_panel}(a,i). In Ref.~\cite{Mackintosh2024}, the V-shaped structure of the sail was chosen based on geometric optics, such that the aberrated light is reflected predominantly to the left, creating a momentum transfer opposing the transverse sail motion. The issue with the V-mirror implementation is that the damping is rather weak: even in the best scenario such a sail is predicted to reduce initial transverse velocities by a mere 33\% over the acceleration phase~\cite{Mackintosh2024}.

Here, rather than a ray optics treatment used for the limited case of the V-mirror~\cite{Mackintosh2024}, we develop a much more general, relativistic wave treatment. This formalism allows us to describe the relativistic optical drag force for arbitrary geometries, and reveals a so-far ignored geometry-dependent term that can be used to enhance the drag force by orders of magnitude.
 
We start in Sec.~\ref{sec:theory} by deriving the generalized Poynting-Robertson forces. As an example of the achievable damping enhancement, we apply our formalism to model a purely reflecting diffraction grating in Sec.~\ref{sec:results}, finding an order-of-magnitude improvement in transverse velocity attenuation compared to the V-mirror~\cite{Mackintosh2024}, and with significantly reduced laser power requirements. We conclude in Sec.~\ref{sec:conclusion} by discussing the significance of the relativistic damping and possible extensions.

%%%
\section{\label{sec:theory}Theory}
%%%

We derive the equations of motion for an arbitrary object irradiated by a plane wave using the framework of the Poynting-Robertson effect. The key extension we make, which is not treated in the standard literature on the Poynting-Robertson effect~\cite{Burns1979,klacka2008}, is that the radiation-pressure cross sections are functions of the angle of plane-wave incidence, as lightsails are far from isotropic. We show that with appropriate sail design this dependence on incident angle can be used to enhance the damping force considerably. 
To get the essence of the effect, we restrict the model to two dimensions (2D) and ignore rotations and thermal effects.

\subsection{Reference Frames}
\begin{figure}[htb]
    \centering
    \includegraphics[width=0.9\linewidth]{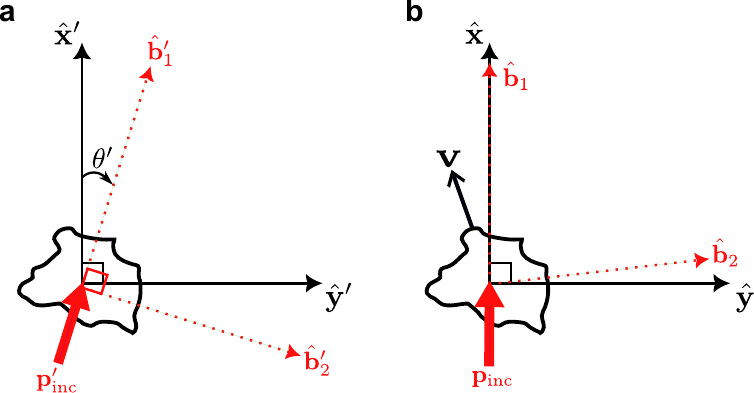}
    \caption{Axes for an arbitrary lightsail (black jagged shape) in the (a) sail and (b) laser frames. (a) Incident laser momentum $\mathbf{p}_\text{inc}'$ is relativistically aberrated relative to $\hat{\mathbf{x}}'$. (b) The orthogonal $\hat{\mathbf{b}}$ axes in (a) are non-orthogonal in the laser frame.}
    \label{fig:axes_definition}
\end{figure}

We require the forces in the frame of the laser on Earth because the ideal sail trajectory is set from this frame and the goal is to minimize transverse perturbations relative to this trajectory. However, radiation forces depending on the local angle of incidence are most easily described in the sail's co-moving frame. We thus define the four-vector orthonormal bases for the laser (\Lframe{}) and instantaneously co-moving sail (\Mframe{}) frames, respectively, as
\begin{align} 
&\Lfourbasis0 \xrightarrow[\text{\Lframe{}}]{} (1,0,0,0),
&&\Lfourbasis j \xrightarrow[\text{\Lframe{}}]{} (0,\Lbasis{j}) \label{eq:L_basis}
\,,
\\
&\Mfourbasis0 \xrightarrow[\text{\Mframe{}}]{} (1,0,0,0), 
&&\Mfourbasis j \xrightarrow[\text{\Mframe{}}]{} (0,\Mbasis{j}') \label{eq:M_basis}
\,.
\end{align}
The four-vector to the left of each arrow has components written in the four-tuple to the right of each arrow, as measured in the frame of reference labeled below the arrow. The index $j\in\{1,2\}$  denotes the spatial direction (and we ignore  $j=3$ components in our 2D model). 
Primed quantities are measured in the sail's comoving reference frame \Mframe{} in which the sail is momentarily at rest.
Frame \Mframe{} has velocity $\bf{v}$ relative to \Lframe{} and corresponding four-velocity $\vec{u}$. 
The spatial basis vector $\Lfourbasis1$ is defined such that its spatial three-vector component $\Lbasis{1}$ points from Earth to the target destination, \textit{e.g.} Proxima Centauri, and defines the ``longitudinal'' direction. For convenience, we redefine the spatial components in more conventional terms: $\Lbasis{1} \equiv \hat{\mathbf{x}}$ and $\Lbasis{2} \equiv \hat{\mathbf{y}}$, where $\hat{\mathbf{y}}$ represents the ``transverse'' direction in which the sail should be asymptotically stable (see Fig.~\ref{fig:axes_definition}). 
The prime on the indices in Eq.~\eqref{eq:M_basis} distinguishes the basis vectors of frame \Mframe{} from frame \Lframe{}. 
The basis vectors in Eq.~\eqref{eq:M_basis} are obtained by inverse Lorentz transformation of those in Eq.~\eqref{eq:L_basis}~\cite{schutz2009}, \textit{i.e.} $\vec{f}_{\mu'} = \Lambda(-\mathbf{v})_{\mu'}^\nu \vec{f}_\nu$ where the Greek indices span $0$ to $3$ and we adopt the Einstein summation convention for lower and upper index pairs. 
 The component $\Lambda(\mathbf{v})_{\mu'}^\nu$ is the $(\nu,\mu')$ element of the Lorentz transformation matrix~\cite{frahm1979},
\begin{equation*} \label{eq:LorentzBoost}
    \Lambda(\mathbf{v})=\left(
    \begin{array}{cccc}
     \gamma  & -\frac{\gamma  v_x}{c} & -\frac{\gamma  v_y}{c} & -\frac{\gamma  v_z}{c} \\
     -\frac{\gamma  v_x}{c} & 1+\frac{(\gamma -1) v_x^2}{v^2} & \frac{(\gamma -1) v_x v_y}{v^2} & \frac{(\gamma -1) v_x v_z}{v^2} \\
     -\frac{\gamma  v_y}{c} & \frac{(\gamma -1) v_x v_y}{v^2} & 1+ \frac{(\gamma -1) v_y^2}{v^2} & \frac{(\gamma -1) v_y v_z}{v^2} \\
     -\frac{\gamma  v_z}{c} & \frac{(\gamma -1) v_x v_z}{v^2} & \frac{(\gamma -1) v_y v_z}{v^2} & 1+ \frac{(\gamma -1) v_z^2}{v^2} \\
    \end{array}
    \right)
    \,,
\end{equation*}
with $v_z = 0$.

We define four-vectors $\vec{b}_j$ as photon four-momenta which, in frame \Mframe{}, have unit time-like components and spatial three-vector components parallel and perpendicular to the beam for $j=1$ and $j=2$, respectively (see Fig.~\ref{fig:axes_definition}(a)). In \Lframe{} (Fig.~\ref{fig:axes_definition}(b)), where the laser points in the acceleration direction, the spatial part of $\vec{b}_1$ points in the $\hat{\mathbf{x}}$ direction. In \Mframe, relativistic aberration creates an angle $\theta'$ between $\hat{\mathbf{x'}}$ and the spatial part of $\vec{b}_1$ as shown in Fig.~\ref{fig:axes_definition}(a).

\subsection{Poynting-Robertson effect and radiation-pressure cross sections}
Our derivation begins with the covariant equation of motion for an arbitrarily shaped particle (sail) in the standard Poynting-Robertson effect. The change in sail four-momentum is derived by subtracting the four-momentum scattered by the sail from the incident four-momentum of a plane wave, yielding 
~\cite{klacka2008,klacka2014}
\begin{equation} \label{eq:general_EOM}
\frac{d\vec{p}}{dt'}
=
\frac{D^2I\Cpr{1}'}{c} \left(\vec{b}_1 - \frac{\vec{u}}{c} \right) 
+ \frac{D^2I\Cpr{2}'}{c} \left(\vec{b}_2 - \frac{\vec{u}}{c} \right)
\,. 
\end{equation}
Here, $d\vec{p}/dt'$ is the derivative of the sail four-momentum with respect to the sail's proper time, $I$ is the laser intensity in the laser rest frame, and $c$ is the speed of light. The factor $D$ is shorthand for the relativistic Doppler factor $D(\mathbf{v}) = \gamma(\mathbf{v}) (1-v_x/c)$, where $\gamma(\mathbf{v}) = [1-v^2/c^2]^{-1/2}$ is the Lorentz factor and $v\equiv \norm{\mathbf{v}}$ is the sail speed relative to the laser.

 The key factors in Eq.~\eqref{eq:general_EOM} are the radiation-pressure cross sections $\Cpr{1}'$ and $\Cpr{2}'$, which are the basis of our theoretical framework.   
The cross sections $\Cpr{j}'$ are defined such that an incoming intensity $I'$ in the sail frame generates a radiation pressure $I'\Cpr{j}'/c$ in the spatial $\vec{b}_j$ direction (see Fig.~\ref{fig:axes_definition}(a)).
As an example, a perfectly reflecting sphere with radius $R'$ has cross sections $\Cpr1' \equiv \pi R'^2$ and $\Cpr2' \equiv 0$ since the cross section over which the particle can reflect light is the geometric cross-sectional area of the sphere (the forces on the sphere due to diffraction are symmetric with respect to the line of plane wave incidence, see the extinction paradox in Ref.~\cite{vdH1981}). In other words, the force on the reflecting sphere is purely in the direction of incident light and depends only on the sphere's geometric cross section. 
We may thus interpret the terms $\vec{b}_{1,2}$ in Eq.~\eqref{eq:general_EOM} as the directions of radiation pressure with magnitude $\Cpr{1,2}'$ and the terms containing $\vec{u}$ as the relativistic aberration drag which is linear and opposite in sign to the four-velocity.

The key modification we introduce is that we do not take $\Cpr{j}'$ to be constants, instead accounting for both incident angle and wavelength dependence: $\Cpr{j}' = \Cpr{j}'(\theta',\lambda')$. In the sail frame, these dependencies arise from the sail's velocity relative to the laser: the incident angle becomes  
the relativistic aberration angle $\theta'$ and the wavelength $\lambda'$ increases due to the relativistic Doppler shift (even at fixed laser wavelength $\lambda$), important for dispersive nanostructured sails. The quantities $\theta'$ and $\lambda'$ encode the effect of the Lorentz transformation (between \Lframe{} and \Mframe{}) on the spatial and temporal light components, respectively.

\subsection{Equation of Motion}
To get the forces on the sail $dp^x/dt'$ and $dp^y/dt'$ in frame \Lframe{}, take the Minkowski dot product of Eq.~\eqref{eq:general_EOM} with $\Lfourbasis1$ and $\Lfourbasis2$, respectively. This procedure results in dot products between $\vec{b}_j \rightarrow_{\text{\Mframe{}}} (1,\hat{\mathbf{b}}_j')$ and $\vec{f}_i$ for $i,j \in \{1,2\}$. The frame-invariant dot products are most easily calculated in frame \Mframe{}, where the $\hat{\mathbf{b}}_j'$ axes are orthogonal and readily expressed in terms of the relativistic aberration angle $\theta'$ as shown in Fig.~\ref{fig:axes_definition}(a). For example, after some algebra, the transverse component of the four-vector $\vec{b}_2$ is found to be
\begin{align*}
\begin{split}
    \Lfourbasis2 \cdot \vec{b}_2
    &=
    \eta_{\mu\xi} f_2'^\xi b_2'^\mu 
    \\
    &=
    \eta_{\mu\xi} \Lambda(\mathbf{v})_\nu^\xi f_2^\nu b_2'^\mu 
    \\
    &=
    - \frac{\gamma v_y}{c}
    - \cos\theta' \left[1 + (\gamma-1)\frac{v_y^2}{v^2} \right] \\
    &\hspace{0.5cm}+ \sin\theta' (\gamma-1) \frac{v_xv_y}{v^2}
    \,,
\end{split}
\end{align*}
where $\eta_{\mu\xi}$ is the $(\mu,\xi)$-component of the Minkowski metric with sign convention $(+1,-1,-1,-1)$. A similar calculation can be performed for $\Lfourbasis1\cdot\vec{b}_2$, and the remaining dot products such as $\Lfourbasis1\cdot\vec{b}_1$ are comparatively straightforward. After evaluating all dot products in terms of velocity components or $\theta'$, we find the longitudinal and transverse equations of motion in \Lframe{}, respectively, to be
\begin{widetext}
\begin{align}
\frac{dp^x}{dt'}
={}&
\frac{D^2I}{c} 
\Bigg\{ 
	\Cpr{1}'(\theta',\lambda') \left[\frac{1}{D} -\frac{\gamma v_x}{c} \right]
	- \Cpr{2}'(\theta',\lambda')
	\left[ 
		\sin\theta' \left(1 + (\gamma-1)\frac{v_x^2}{v^2} \right)
		- \cos\theta' (\gamma-1)\frac{v_xv_y}{v^2}
	\right]
\Bigg\}
\,, \label{eq:dp^x/dt}
\\
\frac{dp^y}{dt'}
={}&
\frac{D^2I}{c} 
\Bigg\{ 
	\Cpr{1}'(\theta',\lambda') \left[ -\frac{\gamma v_y}{c} \right]
	- \Cpr{2}'(\theta',\lambda')
	\left[ 
		\sin\theta' (\gamma-1) \frac{v_xv_y}{v^2} 
            - \cos\theta' \left(1 + (\gamma-1)\frac{v_y^2}{v^2} \right)
	\right]
\Bigg\} \,.
\label{eq:dp^y/dt}
\end{align}
\end{widetext}
In Eqs.~\eqref{eq:dp^x/dt} and~\eqref{eq:dp^y/dt}, $\theta'$ as a function of $v_x$ and $v_y$ is obtained by converting the known laser four-momentum in frame \Lframe{} (proportional to $(1,1,0,0)$) to frame \Mframe{}. We take all expressions to first order in $v_y/v\ll 1$, a reasonable assumption due to the immense longitudinal radiation pressure. Under this assumption, $\gamma$ and $D$ are independent of $v_y$ but a first-order Taylor expansion for $\theta'$ reveals
\begin{align} \label{eq:theta}
\theta' \simeq -\left(\frac{1}{D}-1\right) \frac{v_y}{v} \,.
\end{align}
%.
Finally, we take the Taylor expansion of the radiation-pressure cross sections $\Cpr{j}'$ with respect to $\theta'$ in Eqs.~\eqref{eq:dp^x/dt} and~\eqref{eq:dp^y/dt}, replace $\theta'$ by Eq.~\eqref{eq:theta} and retain linear $v_y/v$ terms. The resulting first-order lightsail equations of motion are then 
\begin{align}
\frac{dp^x}{dt'}
&\simeq
\frac{\gamma D^2I}{c}
\bigg[
    \Cpr{1}'(0,\lambda') 
    + \frac{\Cpr{2}'(0,\lambda')}{D} \frac{v_y}{c} \nonumber\\
    &\hspace{1.6cm}- \Bigl(\frac{1}{D} - 1\Bigr) \pdCprpdtheta{1}(0,\lambda') \frac{v_y}{v}
\bigg]
\label{eq:dp1_expanded} \,,
\\
\frac{dp^y}{dt'}
&\simeq
\frac{D^2I}{c}
\bigg[
    \Cpr{2}'(0,\lambda')
    - \gamma \Cpr{1}'(0,\lambda')\frac{v_y}{c} \nonumber\\
    &\hspace{1.6cm}- \Bigl(\frac{1}{D} - 1\Bigr) \pdCprpdtheta{2}(0,\lambda') \frac{v_y}{v}
\bigg]
\label{eq:dp2_expanded}
\,.
\end{align}
The first terms in each of Eqs.~\eqref{eq:dp1_expanded} and~\eqref{eq:dp2_expanded} are standard radiation pressure terms. The second term of Eq.~\eqref{eq:dp2_expanded}, proportional to $\Cpr{1}'(0,\lambda')$, comes from relativistic aberration and can provide transverse damping through the linear $-v_y$ dependence. This term represents the standard Poynting-Robertson effect when the angular dependence of $\Cpr{j}'$ is ignored -- however, it is bounded because $0\leq\Cpr{1}'\leq2\Cext'$~\cite{klacka2014}, where $\Cext'$ is the cross section of total power the sail removes from the laser. 

The term that does not appear in the standard Poynting-Robertson effect is the new, third term of Eq.~\eqref{eq:dp2_expanded} containing $\pdhfrac{\Cpr{2}'(0,\lambda')}{\theta'}$. This angular derivative has no \textit{a priori} limit and thus can be enhanced in carefully designed sail structures. In the case of a spherical sail where $\Cpr{2}' \equiv 0$ for all $\theta'$, only the $\Cpr{1}'(0,\lambda')$ term provides damping, which is independent of incident angle and is greatly limited by the bound on $\Cpr{1}'(0,\lambda')$. Therefore, we require sails in which $\Cpr{2}'$ does {\em not} vanish for $\theta'\neq 0$: an increase in $\Cpr{2}'$ with $\theta'$ provides a transverse force proportional to the transverse velocity, that is, a drag force.  Designing the angular dependence of $\Cpr{2}'(\theta',\lambda')$ allows us to fully exploit the relativistic aberration and significantly enhance the damping. We note that Eqs.~\eqref{eq:dp1_expanded} and~\eqref{eq:dp2_expanded} applied to the V-mirror lead to the same equations of motion as those derived in Ref.~\cite{Mackintosh2024} using ray-optics-based calculations (Appendix~\ref{app:V_mirror}).

%%%
\section{\label{sec:results}Results}
\begin{figure*}[htb]
    \centering
    \includegraphics[width=0.7\textwidth]{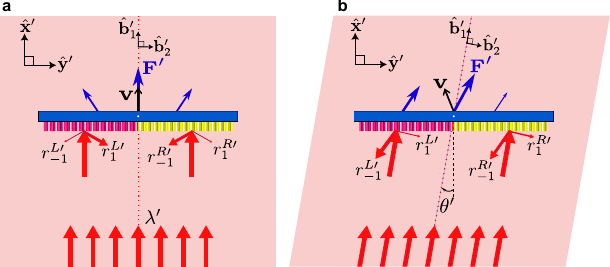}
    \caption{(a) The bigrating in ideal operation with zero transverse velocities and forces. (b) Relativistic aberration due to the leftward velocity results in a change in incident angle $\theta'$ and hence in reflected efficiencies.}
    \label{fig:GRG_damping}
\end{figure*}

%%%
To demonstrate how we can achieve enhanced damping, we model the attenuation of transverse velocities for a metasurface sail consisting of two diffraction gratings called a bigrating. We choose the bigrating because the diffracted orders can be sculpted to provide restoring forces and torques within a Gaussian laser beam~\cite{ilic2019,Srivastava2019}. We take the laser beam to be a plane wave laser with TE (out-of-plane) polarization since we concentrate on the damping force rather than the restoring force. We model the bigrating as in Fig.~\ref{fig:GRG_damping}: a perfectly reflecting, rigid sail consisting of two gratings attached together with mirror symmetry about the $\hat{\mathbf{x}}'$ axis at the center of mass. To keep to the essential physics, we limit ourselves to gratings having only the specular and first diffraction orders, which can be approximated as discrete reflections with reflection coefficients $\effr{m}(\theta',\lambda')$ ($m=0, \pm1$) signifying the normalized power in each order. The angle of diffraction $\theta_m'$ for the $m$-th order is governed by the grating equation $\sin\theta_m' = \sin\theta' + m\lambda'/\Lambda'$, where $\lambda'$ and $\Lambda'$ are the incident wavelength and grating period measured in the grating frame of reference, respectively. The reflection coefficients determine the radiation-pressure cross sections in exact form (derived in Appendix~\ref{app:reflection_grating}) and hence, via Eqs.~\eqref{eq:dp1_expanded} and~\eqref{eq:dp2_expanded}, the grating equations of motion. 

The damping of the bigrating can be understood physically through the non-specular reflection in Fig.~\ref{fig:GRG_damping}. As $\theta'>0$ increases, corresponding to an increase in transverse velocity and relativistic aberration, the orders change direction and power is redistributed. From the grating equation, we can anticipate which orders need to receive the largest share of incoming power to maximize damping. In this case, in both the left and right gratings we require power to flow predominantly into the $m=-1$ order (and out of the other orders) for small leftward transverse velocities. By conservation of momentum, the resultant force on the grating then points predominantly to the right, opposing the transverse velocity. 
A grating for which power is redistributed in this manner has $\pdhfrac{\Cpr{2}'(0,\lambda')}{\theta'}>0$ (see Appendix~\ref{app:reflection_grating}), consistent with the third term in Eq.~\eqref{eq:dp2_expanded} being a drag term. The symmetry of the bigrating with respect to the $\hat{\mathbf{x}}'$ axis at the center of mass ensures that the damping occurs for rightward transverse velocities as well.

To fairly compare the damping performance of different sail designs over the entire acceleration phase, we define a figure of merit by dividing a sail’s damping coefficient by its longitudinal
acceleration (keeping only the main contributing terms in Eqs.~\eqref{eq:dp2_expanded} and~\eqref{eq:dp1_expanded}, respectively). We then average over the relevant wavelength range to produce the figure of merit 
\begin{equation} \label{eq:FOM}
\FOM 
= 
\Braket{
    F_D(\lambda')
    }_{\lambda'}
= 
\Braket{
    \frac{\pdfrac{\Cpr{2}'}{\theta'}(0,\lambda') + \Cpr{1}'(0,\lambda')}{\Cpr{1}'(0,\lambda')} 
    }_{\lambda'}
\,,
\end{equation}
where the angle brackets denote an average of the single-wavelength figure of merit $F_D(\lambda')$ over the Doppler-broadened wavelength range $\lambda' = \lambda$ to $\lambda' = \lambda/D(\mathbf{v}_f)$, corresponding to acceleration from $v=0$ to $v=v_f$. The figure of merit can be interpreted as the ratio of the sail's flight time to the sail's damping time (the time taken for a linearly damped sail to have its initial transverse velocity reduced by a factor $e\approx 2.7$) and depends solely on the sail's optical properties (see Appendix~\ref{app:FOM}).
For a sail with target longitudinal velocity $\beta_f = v_f/c$ and an initial unwanted transverse velocity $v_{y,0}$, the figure of merit can be used to estimate the final transverse velocity by $v_{y,f} \approx v_{y,0} \exp(-\beta_f \FOM)$, with a velocity-averaged (constant) damping coefficient in Eq.~\eqref{eq:dp2_expanded}. We compare sails with the same $v_f$, mass and geometric cross section assuming that the sails have the same acceleration distance, but one may choose to keep other parameters fixed.

For comparison, a perfectly reflecting sphere with any radius has a baseline $(\FOM)_\text{sphere} = 1$. The V-mirror's figure of merit takes the exact form $(\FOM)_\text{VM} = 2\cot^2(\alpha')$ for $\alpha'$ the half angle between the mirrors (Fig.~\ref{fig:VM_panel}(a,i)), derived in Appendix~\ref{app:V_mirror} and consistent with Eq.~(17) in Ref.~\cite{Mackintosh2024}. For the configuration where $\alpha'=\pi/4$, $(\FOM)_\text{VM} = 2$ and the predicted attenuation for $\beta_f = 0.2$ is $1-v_{y,f}/v_{y,0} \approx 33\%$. With vanishing angle between the mirrors the figure of merit diverges, however, this is because the diminishing optical cross section reduces the acceleration. This leaves more time for the damping to be effective (the denominator in Eq.~\eqref{eq:FOM} goes to zero), but comes at the cost of an increase in laser power to reach $v=0.2c$ over a fixed acceleration distance. This trade-off is clearly not advantageous for a lightsail mission.

\begin{figure*}[htb]
    \centering
    \includegraphics[width=0.9\textwidth]{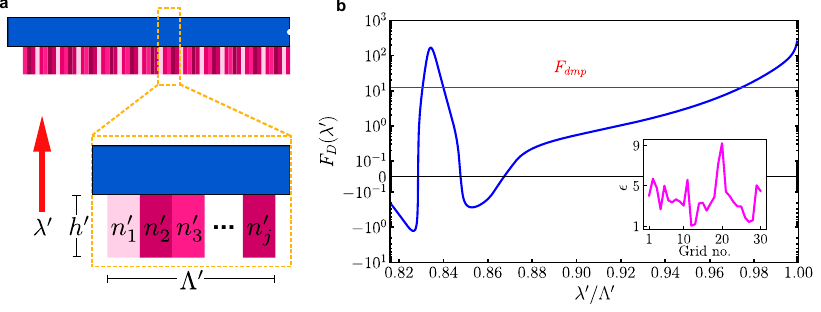}
    \caption{(a) The grating unit cell with period $\Lambda'$ and optimisation parameters ($j=30$ for our grating). (b) Single-wavelength figure of merit $F_D$ (blue) over the normalized, Doppler-shifted wavelength for the optimized grating with permittivity profile shown in the inset (magenta) and $\theta'=0$ plane-wave incidence. The red line is $\FOM$, the average of $F_D$ over the plotted wavelength range.}
    \label{fig:opt_grating}
\end{figure*}

The figure of merit in Eq.~\eqref{eq:FOM} suggests that a grating with broadband resonance in $F_D(\lambda')$ maximizes the damping. The resonance bandwidth ideally covers the approximate wavelength range $[\lambda,1.22\lambda]$, corresponding to the Doppler shift experienced for a sail accelerating from $v=0$ to $v=0.2c$. However, it is difficult to achieve broadband grating resonances over such a wide range. Therefore, we employ inverse design to optimize $\FOM$ with many degrees of freedom in the unit cell, based on the optimization of Jin {\sl et al.} \cite{Jin2020}. In addition to the laser emission wavelength $\lambda$ and grating thickness $h'$, we vary the real, non-dispersive refractive index of $30$ grid points in the unit cell shown in Fig.~\ref{fig:opt_grating}(a) (see Appendix~\ref{app:numerics} for details). The refractive index of each grid is upper bounded by silicon with $n_\text{Si} = 3.5$, in principle allowing fabrication through nanostructuring. To ensure that the grating is entirely reflective (and noting that transmitted orders would only diminish acceleration without changing the damping problem qualitatively), we take the grating to have a lossless reflecting substrate with a large and negative relative permittivity. 

The resulting optimum grating structure is shown in the Fig.~\ref{fig:opt_grating}(b) inset. In Fig.~\ref{fig:opt_grating}(b), we plot $F_D(\lambda')$ for this grating over the full Doppler-broadened laser line starting at $\lamhfrac = 0.816$. The figure shows that the damping force is largest in the wavelength range of the Fano resonance at $\lamhfrac \approx 0.83$ where $F_D(\lambda') > 0$. In this range, $F_D$ is enhanced by a factor $100$ compared to the V-mirror. The wavelengths where $F_D(\lambda') < 0$ correspond to regions of anti-damping (transverse velocity boosting) which are undesirable, however, the boosting is orders of magnitude weaker than the damping (note the logarithmic vertical scale). Once the sail velocity is large enough for the grating to cross the threshold $F_D(\lambda') = 0$ at $\lamhfrac \approx 0.865$, the damping steadily increases until it reaches the first order cutoff at $\lamhfrac = 1$. Over the entire acceleration phase (corresponding to the horizontal axis in Fig.~\ref{fig:opt_grating}(b)), $F_D(\lambda')$ has an average value $\FOM = 12.3$, which predicts a transverse velocity attenuation by $1-v_{y,f}/v_{y,0} = 91.5\%$ for $\beta_f = 0.2$. 

\begin{figure*}[htb]
    \centering
    \includegraphics[width=0.8\textwidth]{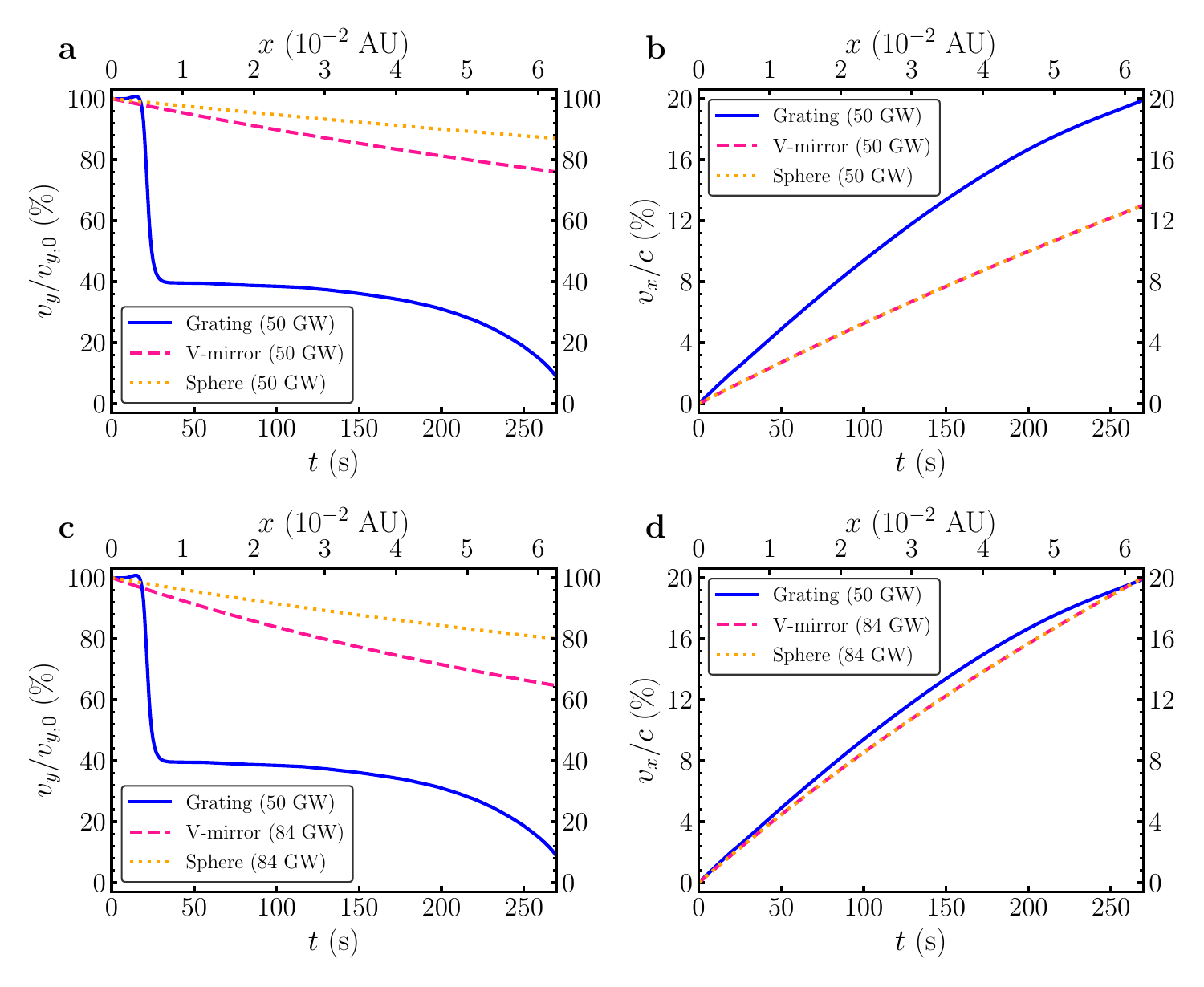}
    \caption{(a) Transverse and (b) longitudinal velocity dynamics for three sails intercepting the same laser power. (c) Transverse and (d) longitudinal velocity dynamics but with the V-mirror and spherical sails intercepting higher laser power.}
    \label{fig:dynamics}
\end{figure*}

We calculated the dynamics of our optimized bigrating by numerically solving Eqs.~\eqref{eq:dp1_expanded} and~\eqref{eq:dp2_expanded} (see Appendix~\ref{app:numerics} for details). We set nominal total sail mass and power intercepted as $m = 1$~g and $P = 50$~GW as per current estimates~\cite{lubin2016}. The laser beam is a plane wave with uniform intensity, so the sail cannot provide a restoring force, allowing us to concentrate on the damping. For this reason, we applied the initial condition of a small transverse velocity $[x,y,v_x,v_y] = [0,0,0,1~\text{m\hspace{0.5mm}s}^{-1}]$ at $t=0$, with the dynamics at later times displayed in Fig.~\ref{fig:dynamics}. The final transverse velocity reduction (Fig.~\ref{fig:dynamics}(a)) for the optimized bigrating is 92.5\%, consistent with the prediction using the figure of merit. The profile of the reduction over time is also consistent with the damping profile in Fig.~\ref{fig:opt_grating}(b). In particular, the resonance region contributes to the sharp decline in $v_y$ at $t \approx 20$~s and the increase in damping force up to cutoff produces the subsequent $v_y$ drop off until the sail reaches $v=0.2c$. Additionally, we see the slight increase in $v_y$ seconds prior to the drop at $t\approx 20$~s corresponding to the anti-damping region for $\lamhfrac \lessapprox 0.83$ in Fig.~\ref{fig:opt_grating}(b). 

For comparison to the bigrating, we derived the equations of motion for the spherical and V-mirror sails using Eqs.~\eqref{eq:dp1_expanded} and~\eqref{eq:dp2_expanded} with the latter's expressions presented in Appendix~\ref{app:V_mirror}. The numerical solution for the sphere and V-mirror ($\alpha'=\pi/4$) dynamics are presented alongside the bigrating in Fig.~\ref{fig:dynamics}. The first row of Fig.~\ref{fig:dynamics} shows the dynamics of all three sails with the same mass, geometric cross section and intercepted power. In this case, the sphere and V-mirror longitudinal accelerations coincide and are incapable of accelerating the sails to $v_f=0.2c$. In the second row of Fig.~\ref{fig:dynamics}, we increased the power intercepted by the sphere and V-mirror so that all sails reach $v_f=0.2c$ with the same flight time. The sphere has a meagre 20.0\% reduction in the initial transverse velocity over the acceleration period, with the V-mirror only marginally better at 35.3\% reduction. These results show that nanostructured sail designs, taking advantage of resonant enhancement, offer a remarkable damping and acceleration improvement compared to mirror-based designs.

%%%
\section{\label{sec:conclusion}Discussion and conclusions}
%%%
Relativistic aberration in the Poynting-Robertson effect can damp residual transverse velocities acquired during the lightsail acceleration phase. The critical addition that to our knowledge has been neglected in the Poynting-Robertson effect literature~\cite{Burns1979,klacka2008} is that, for a lightsail, the optical-cross-section dependence on incident angle needs to be taken into account. This dependence can be designed via the nanostructure to increase damping by orders of magnitude, as illustrated by our bigrating example. However, our derived translational forces can be applied to sail structures with arbitrary scattering properties and are not limited to the bigrating. In principle, the angular derivative of the transverse radiation pressure is unbounded and the damping force can be arbitrarily increased through further optimization. This could involve more adeptly exploiting the broadband grating resonances or transitioning the model to other photonic structures such as metasurfaces which have been shown to exhibit restoring dynamics~\cite{siegel2019,salary2020,Taghavi202204,Taghavi202211}. Regardless of the structure, such optimizations would need to be performed again once the sail materials are known~\cite{atwater2018}.

While our damping method shows great potential, asymptotic stability requires the sail to exhibit a restoring force, necessitating a non-uniform laser beam. However, the damping mechanism discussed here can in principle be derived for such a beam using finite-beam transformations between reference frames~\cite{Yessenov2023}. Moreover, asymptotic stability requires a restoring and damping torque on the sail. We expect that damping torques for photonic structures can be straightforwardly derived using the theoretical machinery developed here. In the rotating regime, the sail receives a position-dependent relativistic Doppler shift, in addition to relativistic aberration. In such cases, a frequency-dependent grating response to take advantage of the Doppler shift could substantially enhance rotational damping, analogous to the angular-dependent grating response discussed here which harnesses the relativistic aberration for translational-damping enhancement.    

Thus, with proper optical design, a lightsail can in principle have restoring \textit{and} damping forces bringing asymptotic stability within its propelling beam, without requiring any additional active elements, damped internal degrees of freedom or time-dependent beam modulation. This greatly simplifies sail design within the limited mass budget, and considerably increase the chances of success for the mission.

%%%
\section{Data availability}
%%%
The supporting data and codes for this article are openly available on GitHub~\footnote{J. Y. Lin, C. M. de Sterke, M. S. Wheatland, A. Y. Song, B. T. Kuhlmey, 2024, \protect\url{https://github.com/jadonylin/lightsail-damping}.}.

% \begin{acknowledgments}

% \end{acknowledgments}

\appendix
% Star when there is only one section

%%%
\section{\label{app:V_mirror}Derivation of V-mirror forces}
%%%
Here, we use our radiation-pressure-cross-section formalism to calculate the forces on the V-mirror and compare to the ray-optics approach in Ref.~\cite{Mackintosh2024} to ensure both approaches yield the same result. Assuming the mirrors are large compared to the laser wavelength so diffraction can be ignored, the V-mirror scatters light into two discrete angles $\zeta_L'$ and $\zeta_R'$, one for each mirror. In other words, the scattering function $F'$, \textit{i.e.} the power distribution over scattering angles $\zeta'$, can be expressed as the weighted sum of Dirac delta functions:
\begin{align} \label{eq:VM_scattering_function}
\begin{split}
    F'(\zeta')
    &=
    \frac{2\pi}{\lambda'} \Big[l'\sin(\alpha' + \theta') \delta(\zeta' - \zeta_L') 
    \\
    &\hspace{1.3cm}+ l'\sin(\alpha' - \theta') \delta(\zeta' - \zeta_R') \Big]
    \,, 
\end{split}
\end{align}
where $l'$ is the length of each mirror in Fig.~\ref{fig:VM_panel}(a). The coefficients of the Delta functions are the cross-sectional areas of the left and right mirrors which determine the distribution of incident power into the two reflected rays. 
By substituting Eq.~\eqref{eq:VM_scattering_function} into the definition of the scattering cross section $\Csca'$ and hence $\Cpr{j}'$ in Refs.~\cite{vdH1981,klacka2008}, we find the following radiation-pressure cross sections for the V-mirror:
\begin{align} 
\begin{split}
    \Cpr{1}'(\theta')
    &=
    2l'
    \bigl(
    	\sin\alpha' \cos\theta' - \sin\alpha' \cos\theta' \cos2\alpha'\cos2\theta' 
        \\
        &\hspace{1.3cm}- \sin\theta' \cos\alpha' \sin2\alpha'\sin2\theta'
    \bigr)
    \,, \label{eq:VM_Cpr1}
\end{split}
    \\
\begin{split}    
    \Cpr{2}'(\theta')
    &=
    2l'
    \bigl(
    	\sin\alpha' \cos\theta' \cos2\alpha' \sin2\theta' 
        \\
        &\hspace{1.3cm}+ \sin\theta' \cos\alpha' \sin2\alpha' \cos2\theta'
    \bigr)
    \,. \label{eq:VM_Cpr2}
\end{split}
\end{align}
If we substitute Eqs.~\eqref{eq:VM_Cpr1} and~\eqref{eq:VM_Cpr2} into Eqs.~\eqref{eq:dp1_expanded} and~\eqref{eq:dp2_expanded}, we find
\begin{widetext}
\begin{align} 
\left(\frac{dp^x}{dt'}\right)_\text{VM}
&\simeq
\frac{\gamma D^2Il'}{c} (4\sin^3\alpha')
\,, \label{eq:VM_dp1}
\\
\left(\frac{dp^y}{dt'}\right)_\text{VM}
&\simeq
-\frac{D^2Il'}{c} \frac{v_y}{v} 
\left[
    \Dinv (4\cos\alpha' \sin2\alpha') 
    - (\gamma-1) (4\sin^3\alpha')
\right]
\,, \label{eq:VM_dp2} 
\end{align}
\end{widetext}
consistent with previous ray-optics based momentum-transfer calculations~\cite{Mackintosh2024}. Observe that Eqs.~\eqref{eq:VM_Cpr1} and~\eqref{eq:VM_Cpr2} substituted into Eq.~\eqref{eq:FOM} gives the V-mirror figure of merit $(\FOM)_\text{VM} = 2\cot^2(\alpha')$ quoted in Sec.~\ref{sec:results}.

%%%
\section{\label{app:reflection_grating}Derivation of reflection grating forces}
%%%
The radiation-pressure cross section for gratings and bigratings used in Sec.~\ref{sec:results} can be derived explicitly in terms of the grating efficiencies. Since the bigrating consists of two gratings attached together, we first consider the single-grating cross sections. 

Similarly to the V-mirror, we assume the grating is much larger than the laser wavelength and thus diffracts light into discrete orders $m$ with exact angles $\theta_m'$ for $m \geq 0$. In this case, the power reflected and transmitted into order $m$ are the diffraction efficiencies $r_m'$ and $t_m'$, respectively. The associated scattering function again consists of Delta functions as follows:
\begin{align} \label{eq:G_scattering_function}
\begin{split}
    F'(\zeta')
    &=
    \frac{2\pi}{\lambda'} \area\cos\theta'
    \sum_{m=-n}^{n}
    \Big[ 
        \effr{m}(\theta',\lambda') \delta(\zeta'-\zeta^{r\prime}_{m})
        \\
        &\hspace{2cm}+ \efft{m}(\theta',\lambda') \delta(\zeta'-\zeta^{t\prime}_{m})
    \Big]
    \,,
\end{split}
\end{align}
for $\area$ the total single-grating length. The scattering angles $\zeta^{r,t\prime}_m$ are functions of the diffracted angles $\theta_m'$ and are thus determined by the grating equation. This scattering function assumes both reflection and transmission for up to $n$ non-evanescent orders, whose coefficients $\effr{m}$ and $\efft{m}$ are calculated numerically using Maxwell's equations. Since neither $n\geq2$ nor transmitted orders add qualitatively to the physics, we restrict to $n=1$ and assume zero transmission by attaching a reflective, lossless substrate to the grating. Following the procedure for the V-mirror, we find
\begin{align} 
\Cpr{1}'(\theta',\lambda')
&=
\area\cos\theta'
\sum_{m=-1}^1 
\effr{m}(\theta',\lambda')[1+\cos(\theta_m'+\theta')]
\,, \label{eq:RG_Cpr1}
\\
\Cpr{2}'(\theta',\lambda')
&=
-\area\cos\theta'
\sum_{m=-1}^1 
\effr{m}(\theta',\lambda')\sin(\theta_m'+\theta')
\,. \label{eq:RG_Cpr2}
\end{align}

Substituting Eqs.~\eqref{eq:RG_Cpr1} and~\eqref{eq:RG_Cpr2} into Eqs.~\eqref{eq:dp1_expanded} and~\eqref{eq:dp2_expanded} gives the forces on a reflection grating in frame \Lframe{}. The force on a bigrating is the sum of forces on the left and right gratings, each having different efficiencies $\effrL{m}(\theta',\lambda')$ and $\effrR{m}(\theta',\lambda')$. Therefore, the bigrating forces will contain 6 efficiencies in total and their derivatives. However, due to the mirror symmetry of the bigrating about its center, we can convert the left-grating efficiencies to the right-grating efficiencies using the relations
\begin{align} \label{eq:REST_mirror_symmetries}
\effrL{m}(0,\lambda') = \effrR{-m}(0,\lambda'),
\quad
\pdfrac{\effrL{m}}{\theta'}(0,\lambda') = -\pdfrac{\effrR{-m}}{\theta'}(0,\lambda')
\,,
\end{align}
for all $m \geq 0$.
We can further reduce the number of efficiencies needed for a full description using energy conservation: 
\begin{align} \label{eq:energy_cons}
\sum_{m=-1}^1 \effr{m}(\theta',\lambda') = 1 \,,     
\end{align}
and the following relation which is a consequence of the reciprocity theorem~\cite{Maystre2014}
\begin{align} \label{eq:reciprocity}
\effr{0}(\theta',\lambda') = \effr{0}(-\theta',\lambda')
\,,
\end{align}
both valid for lossless single gratings. 
Finally, the diffraction angles $\theta_m'$ in Eqs.~\eqref{eq:RG_Cpr1} and~\eqref{eq:RG_Cpr2} at $\theta'=0$ can be written explicitly using the grating equation, yielding
\begin{align} 
\sin\theta_m'(0,\lambda') = \mlamfrac 
\,,  \\
\cos\theta_m'(0,\lambda') = \sqrtmlamfrac 
\,, \\
\pdfrac{\theta_m'}{\theta'}(0,\lambda') = \frac{1}{\cos\theta_m'(0,\lambda')}
\,. \label{eq:diffraction_angles}
\end{align}
By combining Eqs.~\eqref{eq:RG_Cpr1} to~\eqref{eq:diffraction_angles}, we find the relevant radiation-pressure cross sections at $\theta'=0$:
\begin{widetext}
\begin{align}
\Cpr{1}^{L\prime}(0,\lambda') + \Cpr{1}^{R\prime}(0,\lambda')
&=
4\area\effrR{0}(0,\lambda') + 2\area\big[\effrR{-1}(0,\lambda') + \effrR{1}(0,\lambda')\big] \left[ 1 + \sqrtlamfrac \right]
\,, \label{eq:Cpr1L+R}
\\
\pdfrac{\Cpr{2}^{L\prime}}{\theta'}(0,\lambda') + \pdfrac{\Cpr{2}^{R\prime}}{\theta'}(0,\lambda')
&=
- [\Cpr{1}^{L\prime}(0,\lambda') + \Cpr{1}^{R\prime}(0,\lambda')]
+ 4\area\lamfrac \pdfrac{\effrR{-1}}{\theta'}(0,\lambda') 
\,, \label{eq:PDCpr2L+R}
\end{align}
\end{widetext}
which lead to the bigrating forces upon substitution into Eqs.~\eqref{eq:dp1_expanded} and~\eqref{eq:dp2_expanded}. In particular, Eq.~\eqref{eq:PDCpr2L+R} is the primary contributor to bigrating damping. With our definition of angles, the angular term $\pdhfrac{\effrR{-1}(0,\lambda')}{\theta'}>0$ qualitatively shows that power flow into the $m=-1$ order with a small increase in $\theta'$ from zero leads to damping. This angular derivative term has great potential for resonant enhancement over a finite wavelength band using optimized gratings, as demonstrated in Sec.~\ref{sec:results}.

%%%
\section{\label{app:FOM}Figure of merit interpretation}
%%%
This section provides some intuition on the figure of merit $\FOM$ based on an approximation using Eqs.~\eqref{eq:dp1_expanded} and~\eqref{eq:dp2_expanded}. We interpret the figure of merit physically as the sail flight time $t_f$ divided by the sail damping time $t_d$, which can be understood as follows. Assuming a constant longitudinal acceleration and $v\ll c$, the flight time is approximately
\begin{align} \label{eq:t_f}
t_f
\simeq 
\frac{mv_f}{\braket{dp^x/dt}_{\lambda'}} 
\simeq
\frac{mv_fc}{\braket{D^2 I \Cpr{1}'(\lambda')}_{\lambda'}} 
\,.
\end{align}
The radiation-pressure cross sections are evaluated at $\theta'=0$ incidence. We assume a linear damping force $dp^y/dt \simeq -\xi v_y$ ($\xi$ constant), where, using Eq.~\eqref{eq:dp2_expanded},
\begin{align} \label{t_damp}
\xi 
\simeq \frac{1}{\gamma} 
\Braket{ \frac{D^2I}{c} \left[\pdfrac{\Cpr{2}'}{\theta'}(\lambda') \Dinv \frac{1}{v} + \Cpr{1}'(\lambda') \frac{\gamma}{c} \right] }_{\lambda'}
\,.
\end{align}
The corresponding damping time, the time for which $v_y(t_d) = v_{y,0}/e$, is $t_d=\gamma m/\xi$.

For $v\ll c$, we have $\gamma \approx 1$ and $1/D-1 \approx \beta$, so taking the ratio $t_f/t_d$ reveals that, to reasonable approximation,
\begin{align} \label{eq:FOM_physical}
\FOM \approx \frac{1}{\beta_f} \frac{t_f}{t_d}
\,.
\end{align}
For a given final velocity, $\FOM$ predicts the total transverse velocity attenuation $v_{y,f}/v_{y,0} \approx \exp(-t_f/t_d) \approx \exp(-\beta_f \FOM)$.
Since $\FOM$ depends solely on the sail's optical properties, a given nanostructured sail has a fixed figure of merit. Therefore, such a sail has a fixed transverse velocity attenuation: over a fixed acceleration distance, attempting to increase the flight time by using a less powerful laser will also increase the damping time such that $\FOM$ remains fixed. 

We note that other figures of merit can be crafted from the radiation-pressure cross sections. For example, a figure of merit defined by the product rather than ratio of terms in Eq.~\eqref{eq:FOM} is proportioanl to $1/(t_ft_d)$ which balances longitudinal acceleration against transverse damping. This is a useful figure of merit when damping comes at the cost of longitudinal acceleration, as is the case for, say, the V-mirror with small angle mirrors or transmissive gratings. However, such a figure of merit is less useful for purely reflective gratings where $\Cpr1'$ has a lower bound. Our choice $\FOM$ is appropriate for maximising the transverse velocity attenuation of reflective sails with purely translational motion, but eventually the figure of merit must account for translations and rotations, both with restoring and damping dynamics.

%%%
\section{\label{app:numerics}Numerical simulations}
%%%
The electromagnetic simulations on the dielectric grating shown in Fig.~\ref{fig:opt_grating} were performed using the \texttt{GRCWA} module~\cite{Jin2020} in Python, in which we discretized the unit cell into 30 grid points. We varied the following parameters in the stated ranges (with lengths normalized to the grating period $\Lambda'$): laser emission wavelength in $[0.5,D(\mathbf{v}_f)]$, grating thickness in $[0,1]$ and refractive index of every grid point in $[1,3.5]$. The choice of wavelength-to-period ratio ensures that the grating does not diffract into the $|m|\geq2$ orders nor reaches the first order cutoff accounting for the Doppler shift. The grating thickness is limited to approximately the grating period due to the low-mass-sail restriction and density of silicon-like materials~\cite{atwater2018}. A lossless substrate with relative permittivity $\epsilon_\text{sub} = -10^6$ was attached to the grating to prohibit transmission. The laser was assumed to be monochromatic and TE-polarized. To optimize the figure of merit in Eq.~\eqref{eq:FOM}, we used the multistart Multi-Level-Single-Linkage (MLSL) global optimizer~\cite{RinnooyKan1987pt2} with the internal gradient-based Method of Moving Asymptotes (MMA) local optimizer~\cite{Svanberg1987}, both available with the \texttt{NLopt}~\cite{Johnson2007} Python package. 

The equations of motion in Eqs.~\eqref{eq:dp1_expanded} and~\eqref{eq:dp2_expanded} can be converted from momentum time derivatives to velocity time derivatives using the relativistic momentum equation $p^i = \gamma m v_i$ ($i\in\{x,y\}$) for $m$ the mass of the sail. We assume zero initial longitudinal velocity and $v_{y,0} = 1~\text{ms}^{-1}$ initial transverse velocity. All transverse velocity results scale linearly with $v_{y,0}$, provided the transverse velocity remains much smaller than the total velocity of the sail. Other than the initial transverse velocity, we assume there is no transverse perturbation during the acceleration phase. We solved the equations of motion numerically using a fourth-order-Runge-Kutta method in \texttt{scipy}~\cite{Virtanen2020}, stopping when the sail velocity reached $v=0.2c$.

% \bibliography{references}% Produces the bibliography via BibTeX.


\end{document}